\pdfoutput=1
\RequirePackage{fix-cm}
\documentclass[smallextended]{svjour3}       % onecolumn (second format)
\smartqed  % flush right qed marks, e.g. at end of proof
\usepackage{graphicx}
\begin{document}

\title{Coherent States and  Modified de Broglie-Bohm Complex Quantum Trajectories%\thanks{Grants or other notes
%about the article that should go on the front page should be
%placed here. General acknowledgments should be placed at the end of the article.}
}
%\subtitle{Do you have a subtitle?\\ If so, write it here}

%\titlerunning{Short form of title}        % if too long for running head

\author{Moncy V. John         \and
        Kiran Mathew  %etc.
}

%\authorrunning{Short form of author list} % if too long for running head

\institute{Moncy V. John \at
              Department of Physics, St. Thomas College, Kozhencherri - 689641, Kerala, 
India  \\
              Tel.: +91 9447059964\\
            %  Fax: +123-45-678910\\
              \email{moncyjohn@yahoo.co.uk}           %  \\
%             \emph{Present address:} of F. Author  %  if needed
           \and
           Kiran Mathew \at
              Department of Physics, St. Thomas College, Kozhencherri - 689641, Kerala, 
India
\email{kiran007x@yahoo.co.in}
}

\date{Received: date / Accepted: date}
% The correct dates will be entered by the editor

\maketitle

\begin{abstract} This paper examines the nature of classical correspondence in the case of coherent states at the level of quantum trajectories. We first show that  for a  harmonic oscillator,  the coherent state  complex quantum trajectories  and the complex classical trajectories  are  identical to each other. This congruence in the complex plane, not restricted to high quantum numbers alone, illustrates that the harmonic oscillator in a  coherent state executes  classical motion.  The quantum trajectories we consider are those conceived  in a modified de Broglie-Bohm  scheme. Though quantum trajectory representations  are widely discussed in recent years, identical  classical and quantum trajectories for coherent states are  obtained  only in the present  approach. We may note that this result  for standard harmonic oscillator coherent states is not totally unexpected because of their holomorphic nature. The study is extended to coherent states of a particle in an infinite potential well and  that in a symmetric Poschl-Teller potential by solving for the trajectories numerically.  For the Gazeau-Klauder coherent state of the infinite potential well,   almost identical classical and quantum trajectories are obtained whereas  for the Poschl-Teller potential, though classical trajectories are not regained,  a periodic motion results as $t\rightarrow \infty$. Similar features were found for the SUSY quantum mechanics-based coherent states of the Poschl-Teller potential too, but this time the pattern of complex trajectories is quite different from that of the previous case. Thus we find that the method is a potential tool  in analyzing the  properties of generalized coherent states.

% \subclass{MSC code1 \and MSC code2 \and more}
\end{abstract}

\section{Introduction}

Quantum trajectories, such as those due to de Broglie and Bohm (dBB), Floyd, Faraggi and Matone (FFM), etc., have gained wide attention  \cite{db1,db2,valentini,dBB,holland,carroll,wyattbook,pratim,floyd,faraggi} recently. 
In another attempt, complex quantum trajectories were  conceived    by a modified de Broglie-Bohm (MdBB) approach to quantum mechanics \cite{mvj1,yang1,yang2,yang3,goldfarb,wyatt1,wyatt2,sanz1,sanz2,goldfarb2,mvj2,mvj3}.  These trajectories are obtained by putting  $\Psi \equiv \exp(iS/\hbar)$ in the Schrodinger equation, which results in  an equation similar to the classical Hamilton-Jacobi equation,  in terms of the generally complex function $S(x,t)$. 
Instead of using the Hamilton-Jacobi formalism (where one must use the Jacobi's theorem to obtain trajectories, as in classical mechanics and the FFM trajectory representation \cite{floyd,faraggi}),  an equation of motion \cite{mvj1}

\begin{equation}
\dot{x}=\frac {1}{m}\nabla S, \label{eq:eqmotion}
\end{equation}
is used, where $x$ now is a complex variable $x\equiv x_r+ix_i$. This equation appears similar to but is not the same as that used by de Broglie \cite{db1,db2,valentini} to obtain a velocity field. On integration, it gives   complex  trajectories, in contrast to the real trajectories in the dBB approach. 

Our attempt in this paper is to find the nature of classical correspondence at the level of   quantum trajectories  of particles in  coherent states   \cite{klauderbook,antoinebook}  (denoted as $\phi_z(x,t)$, where $z$ is a complex parameter), moving in potentials  such as the harmonic oscillator,  infinite potential well, a symmetric Poschl-Teller  potential etc. It is found that  the MdBB representation has the  feature of identical  classical and quantum trajectories for harmonic oscillator coherent states.  For the coherent state  of the infinite potential well, one can observe   this feature  in the limit of high $|z|$. But for the Poschl-Teller potential,   such a conclusion cannot be made. However,  periodic motion for the particle results as $t\rightarrow \infty$ for both the Gazeau-Klauder coherent states and the supersymmetric quantum mechanics (SUSYQM)-based coherent states. 

Since the MdBB quantum trajectories are in a complex plane, the demonstration of identical  quantum and classical trajectories is to be made in the complex plane.  The theory of coherent states is inextricably connected to the complex  parameter $z$ and hence  identical  trajectories for the quantum and classical cases in the complex space is not unnatural.

 Given below, in Fig. (1),   are the trajectories for the quantum harmonic oscillator in the lowest energy eigenstates with $n=$ 0, 1, 2, 3 and 4, drawn using  numerical methods. Similarly, in Fig. 2,  the corresponding complex quantum trajectories for a particle in an infinite potential well are drawn numerically. Such solutions for these stationary states  agree very well with the analytical solutions in \cite{mvj1,yang3}. They serve as  benchmark for the numerical method and encourages us to apply it to other problems with  known wave functions as well.

\begin{figure} [ht]
\centering{\resizebox {1.0 \textwidth} {0.1 \textheight }  
{\includegraphics {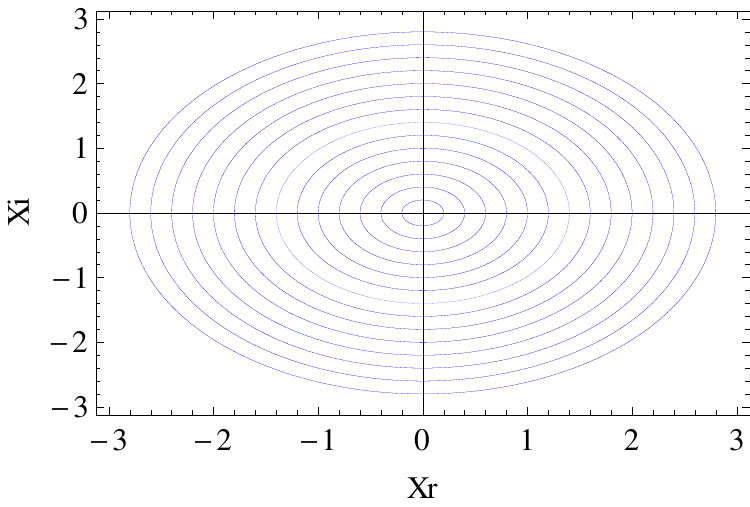}  \includegraphics {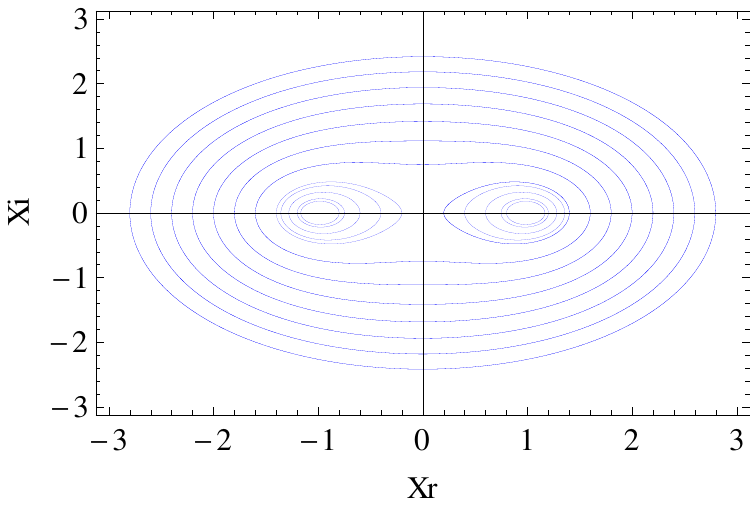} \includegraphics {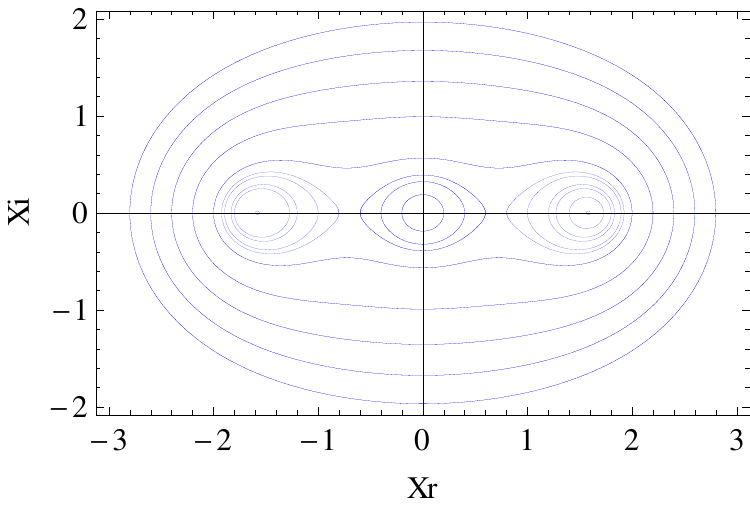} \includegraphics {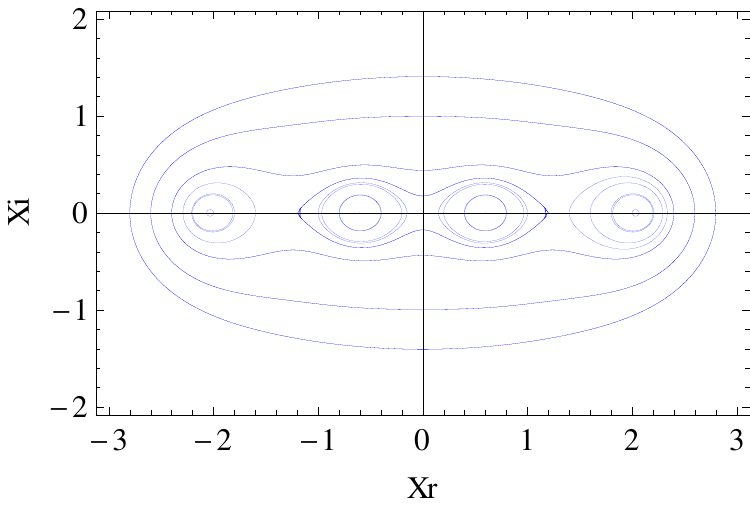} \includegraphics {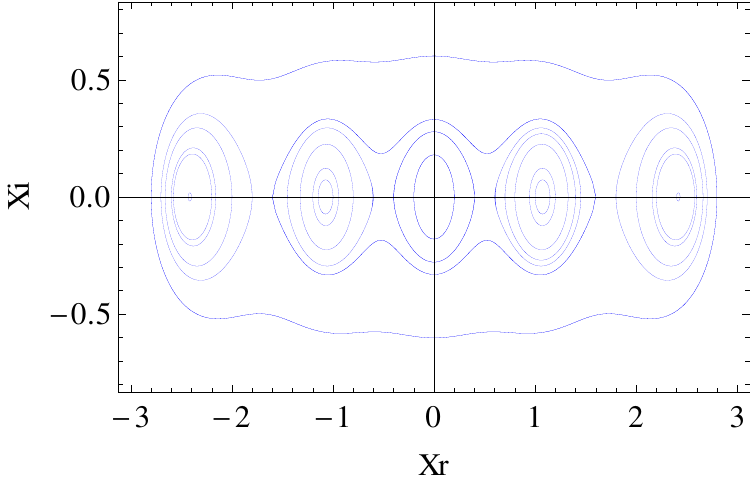}} 
\caption{ The complex quantum trajectories for the  harmonic oscillator in the $n=$0,1,2,3,4 energy eigenstates, respectively, for evenly distributed starting points along the real line. }  \label{fig:ho_qm01234}} 
 \end{figure}

\begin{figure} [ht]
\centering{\resizebox {1.0 \textwidth} {0.1 \textheight }  
{\includegraphics {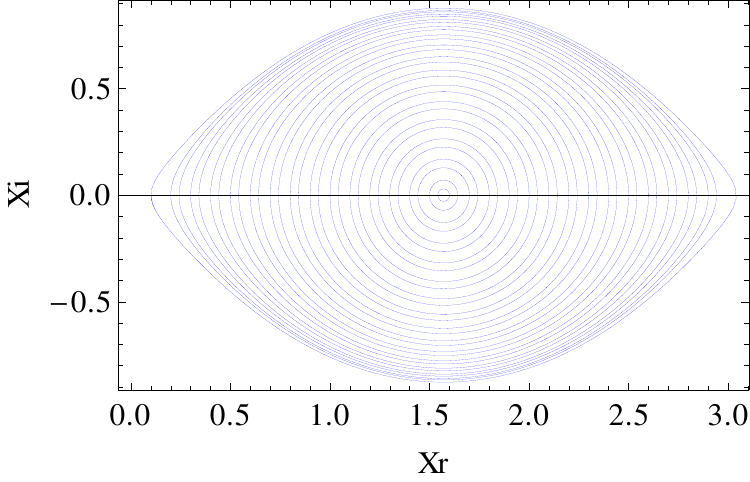}  \includegraphics {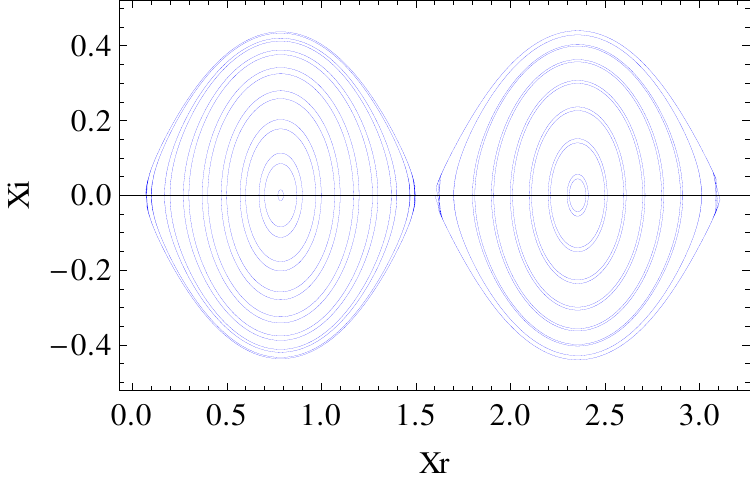} \includegraphics {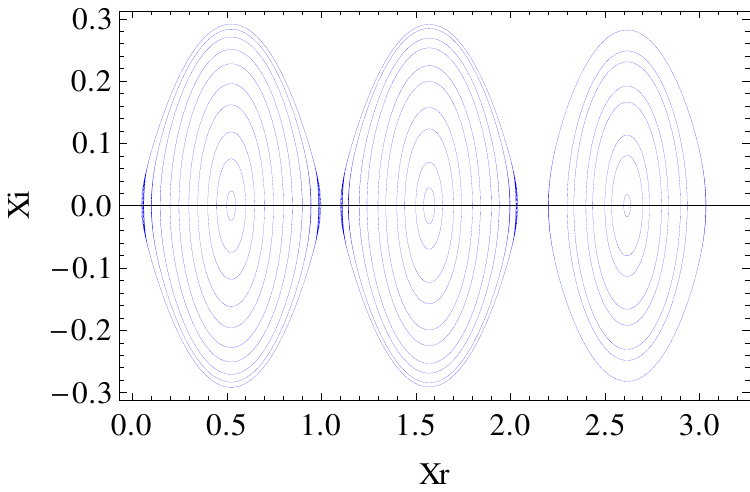} \includegraphics {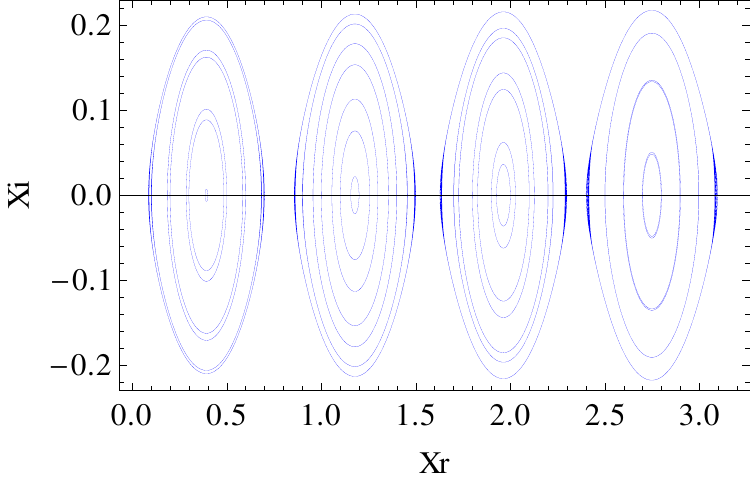} } 
\caption{ The complex quantum trajectories for the particle in an infinite well potential in the $n=$0,1,2,3 energy eigenstates, respectively, for evenly distributed starting points along the real line. }  \label{fig:infwel_qm01234}} 
 \end{figure}

\section{Complex Classical Trajectories}

\begin{figure} [ht]
\centering{\resizebox {0.5 \textwidth} {0.3 \textheight }  
{\includegraphics {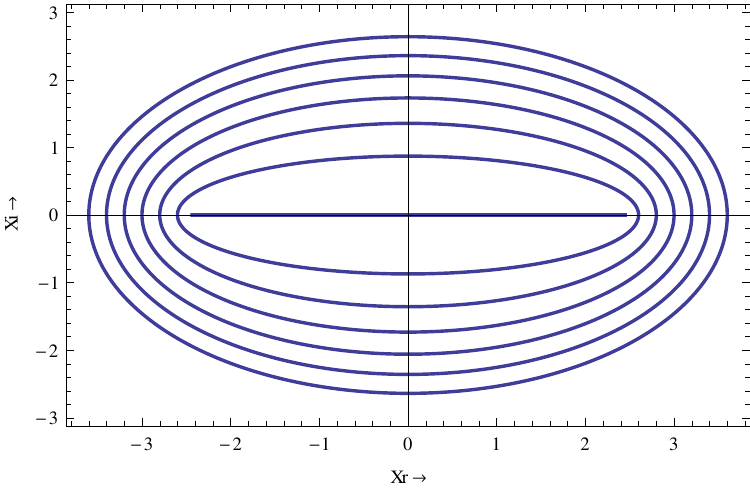} } 
%{\includegraphics {classical_ho.png} } 
\caption{The complex classical trajectories for the  harmonic oscillator of energy $E=4.5$ units. }  \label{fig:ho_cl}} 
 \end{figure}

Solutions of classical dynamical problems of physical systems obtained in terms of complex space variables are  well-known.  In the past decade, interesting properties of  classical trajectories of complex Hamiltonians are discovered \cite{classic1,classic2a,classic3}. To obtain complex trajectories, one solves the Hamilton's equations $\dot{x}=\frac{\partial H}{\partial p}$, $\dot{p}=-\frac{\partial H}{\partial x}$, for complex initial conditions and not just for initial conditions in the classically allowed regions. For instance, consider the harmonic oscillator problem.
With $H=p^2/2m +(1/2)kx^2$, $x\equiv x_r+ix_i$, and  the complex momentum  $p\equiv p_r+ip_i$, one writes the Hamilton's equations  as $\dot{x}_r =p_r/m$, $\dot{x}_i=p_i/m$, $\dot{p}_r=-kx_r$ and $\dot{p}_i=-kx_i$.   For a particle with real energy $E$ such that  $0<E<(1/2)kA^2=(1/2)m\omega^2A^2$, the solution of these equations can be found as

\begin{equation}
x_r=A\cos(\omega t), \qquad x_i=B\sin(\omega t), \qquad B=\sqrt{ A^2 -\frac{2E}{m\omega^2}}. \label{eq:B}
\end{equation}
This leads to elliptical trajectories  in the complex plane, as plotted  in Fig. (\ref{fig:ho_cl}).  Note that in the limit $E\rightarrow 0$, these ellipses become concentric circles.

Now consider the case of a free particle with real energy $E>0$. Then we have $\dot{x}_r=p_r/m$,  $\dot{x}_i=p_i/m$, $\dot{p}_r=0$, $\dot{p}_i=0$, $p_rp_i=0$ and also $p_r^2 >p_i^2$. Then the solutions are $x_r=\pm \sqrt{\frac{2E}{m}} t + c_r$, $x_i=c_i$, where the constants $c_r$, $c_i$ are real. These classical trajectories are straight lines which lie in the complex plane, parallel to the real axis.

\begin{figure} [ht]
\centering{\resizebox {0.5 \textwidth} {0.3 \textheight }  
{\includegraphics {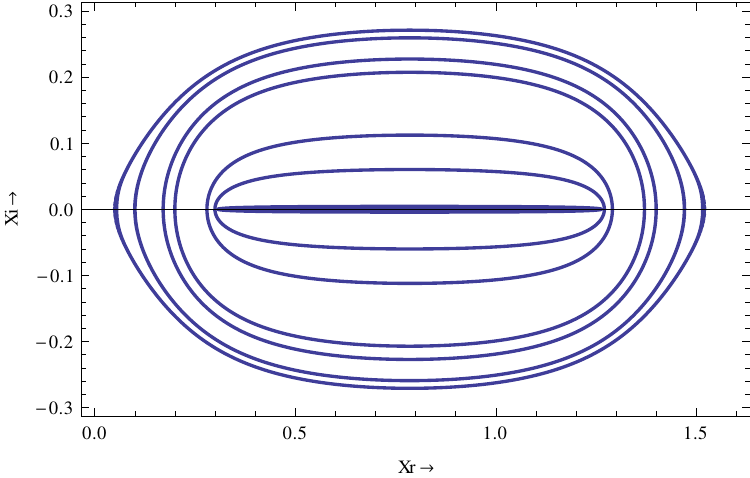} } 
%{\includegraphics {classical_sin2x.png} }
%\includegraphics[width=.4\textwidth]{classical_sin2x.jpg} 
\caption{ The complex classical trajectories for a particle with $E=2.25$ units in the  symmetric Poschl-Teller potential. }  \label{fig:PT_cl}} 
 \end{figure}

Another example we consider here is that of a  Poschl-Teller potential

\begin{equation}
V(x)=\frac{V_0}{2}\left[ \frac{l (l -1)}{\cos^2(\frac{x}{2a})} +\frac{k (k -1)}{\sin^2(\frac{x}{2a})}\right]. \label{eq:PT_pot}
\end{equation}
The symmetric Poschl-Teller potential results when  we take $l =k \geq 1$ \cite{antoinepap}.  Putting $2a$  to be of  unit magnitude, the potential becomes $V(x) =2V_0 l (l -1)/\sin^2(2x)$. The complex classical trajectories for a particle having an energy $E=2.25$ units are drawn  numerically   in Fig. (\ref{fig:PT_cl}).

\section{Complex Quantum Trajectories of Harmonic Oscillator Coherent States}
We may now illustrate that the complex quantum trajectories of the harmonic oscillator in coherent state  are  the same as those trajectories obtained in complex classical mechanics.   Here it is  shown that these  are identical to each other for all mean values of energy, thus demonstrating the quantum-classical correspondence. We start with the coherent state wave function of a harmonic oscillator with $X \equiv \alpha x$,

\begin{equation}
\phi_{z}(x,t)=\left(\frac{\alpha }{\sqrt{\pi}}\right)^{1/2}\exp {\left[\frac{1}{2}\left(X^2-\lambda^2-i\omega t\right)\right]} \exp { \left[ -(X -\eta)^2\right]}, \label{eq:wavefn}
\end{equation}
where $\alpha =\sqrt{m\omega /\hbar}$, $z =\lambda e^{i\kappa}$, and $\eta =\frac{1}{\sqrt{2}}\lambda e^{ -i(\omega t-\kappa)} $. The de Broglie-type equation of motion used in MdBB \cite{mvj1} gives

\begin{equation}
\dot{X} = i\omega (X -2\eta). \label{eq:eqnmotn}
\end{equation}
The above coherent state $\phi_{z}$ is the eigenfunction of the lowering operator $a$ with eigenvalue $z$, which is complex. Note that in the limit $\lambda \rightarrow 0$, the coherent state wave function and its equation of motion reduce to the corresponding entities of harmonic oscillator ground state, as expected. The real and imaginary parts of the velocity field $\dot{X}$ are

\begin{equation}
\dot{X}_r =-\omega \left[ X_i +\sqrt{2}\lambda \sin(\omega t-\kappa)\right] \label{eq:X_rdot}
\end{equation}
and

\begin{equation}
\dot{X}_i=\omega \left[ X_r -\sqrt{2}\lambda \cos (\omega t -\kappa)\right]. \label{eq:X_idot}
\end{equation}
These equations have the  solutions $X_r =A\cos(\omega t-\kappa)$ and $X_i=B \sin(\omega t-\kappa)$, which are the same as the solutions  of the classical Hamilton's equations obtained in (\ref{eq:B}), except for a phase factor, when $A$, $B$ (both real and positive) are related to $\lambda$ by the equation $
A-B=\sqrt{2}\lambda$. Each one of such quantum trajectories in MdBB corresponds to a trajectory  of the complex classical oscillator.  This congruence establishes the correspondence of a quantum harmonic oscillator in coherent state  with the classical harmonic oscillator.  In the limit $\lambda \rightarrow 0$, the solution with $A=B$ exists. This gives concentric circular paths corresponding to the ground state harmonic oscillator obtained in \cite{mvj1,yang3}.

Let us now compare the coherent state MdBB trajectories for the harmonic oscillator with those  in the dBB guiding wave mechanics.  The equation of motion in the latter case is $\dot{X}=-\omega \sqrt{2} \lambda \sin(\omega t-\kappa)$, with $X$ real. The  trajectories are noncrossing \cite{holland} and the particles oscillate with  amplitude $A=\sqrt{2}\lambda$.  In the limit $\lambda \rightarrow 0$ (which is equivalent to the condition that the mean value of energy $E\rightarrow 0$), they remain stationary at  their initial positions $X(0)$. We note that in the real case, classical simple harmonic oscillators can, at best, be stationary only at the equilibrium point $X=0$ and that the above feature of stationary particles at all values of $X$ is unnatural in the classical limit. On the other hand, the MdBB trajectories in the coherent state correspond to the complex classical trajectories in every respect.  Even for $\lambda \rightarrow 0$,  all the trajectories  enclose $X=0$; they are concentric circles of the $n=0$ harmonic oscillator state given in Fig. (\ref{fig:ho_qm01234}) and agree with the corresponding classical trajectories. Often in conventional dBB scheme, the stationarity  of particles in  bound eigenstates  is  justified as an instance which demands  a new `quantum intuition', but the above mentioned failure to exhibit correspondence with classical motion, even for coherent states, is indeed a  setback for the formalism.

\begin{figure} [ht]
\centering{\resizebox {0.5 \textwidth} {0.3 \textheight }  
{\includegraphics {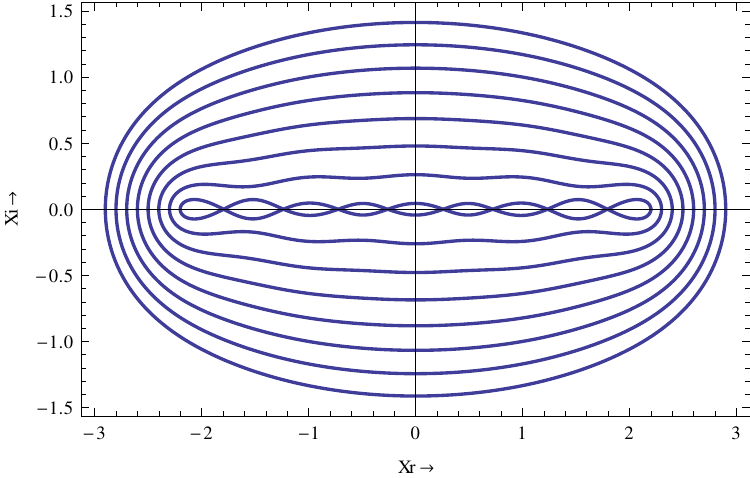} } 
\caption{ The complex trajectories for the harmonic oscillator in a coherent state with $\lambda$ =2.1 and $\kappa=0$, evaluated numerically by including terms up to $n=4$ in the expansion. The trajectories are plotted for the initial values $x_r$ = 2.2, 2.3, 2.4, 2.5, 2.6, 2.7, 2.8 and 2.9}  \label{fig:ho_coh3}} 
 \end{figure}

By using numerical methods the complex trajectories for the harmonic oscillator in a coherent state  with $|z| = \lambda = 2.1$ are drawn. Instead of expression (\ref{eq:wavefn}), we use the expansion 

\begin{equation}
\phi_{z}(x,t)=\sum_{n\geq 0}e^{-\frac{1}{2}\mid z \mid^2} \frac{z^n}{\sqrt{n!}}e^{-i(n+\frac{1}{2})\omega t} \psi_n(x).
\end{equation}
for obtaining the coherent states in terms of the eigenfunctions. The trajectories are shown in Fig. (\ref{fig:ho_coh3}). In this numerical evaluation, we have included terms up to $n=4$ in the expansion. One  finds  deviations   from the expected classical trajectories   at the inner parts. It can easily be seen that  those trajectories with large  values of $x(0)$ are  almost circular in shape and they always agree with the  complex classical trajectories. When  the number of  terms included in the series is varied, however, the shape of  trajectories at the inner regions, with small  values of $x(0)$, vary significantly.   The deviations of these trajectories  with those drawn using Eqs. (\ref{eq:X_rdot}) and (\ref{eq:X_idot}) are thus expected to arise from the truncation in the series. 

\section{Infinite Potential Well and Poschl-Teller  Potential}
\begin{figure} [ht]
\centering{\resizebox {1.0 \textwidth} {0.2 \textheight }  
{\includegraphics {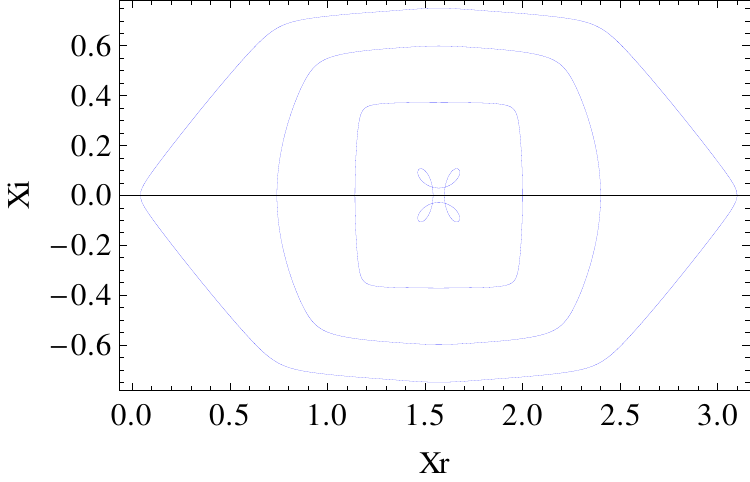}  \includegraphics {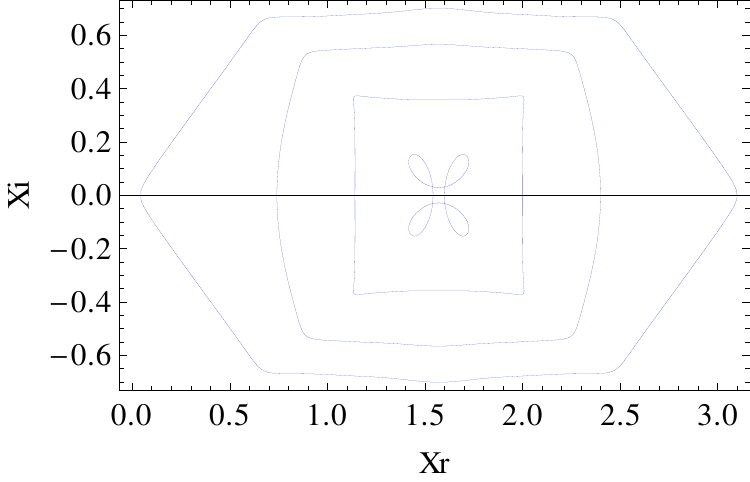} \includegraphics {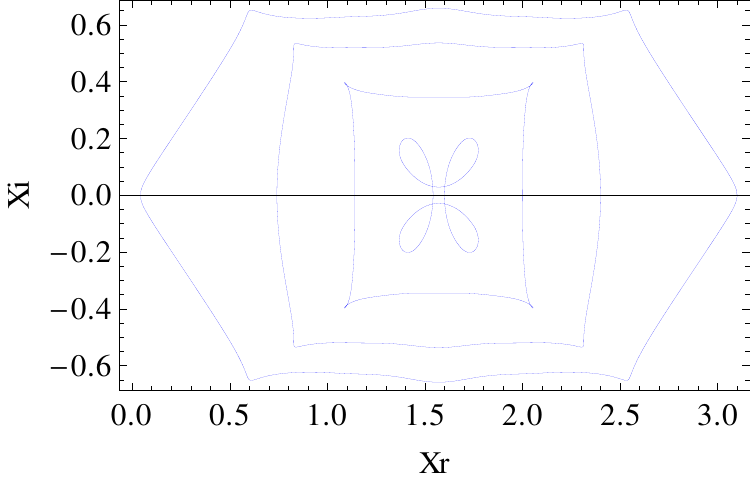} \includegraphics {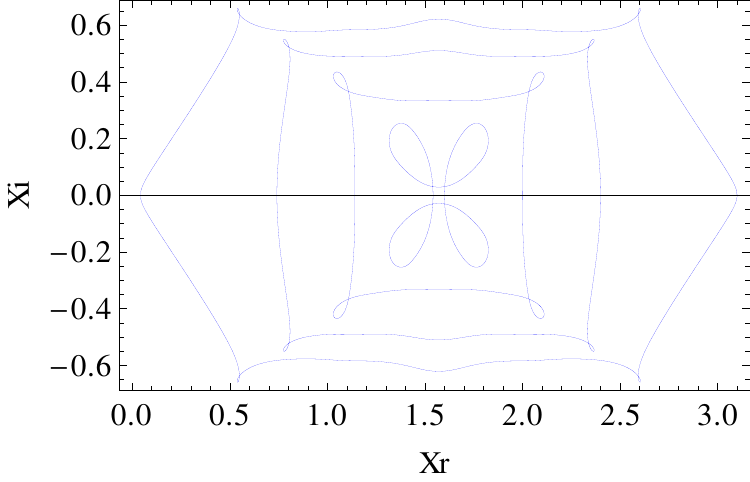} \includegraphics {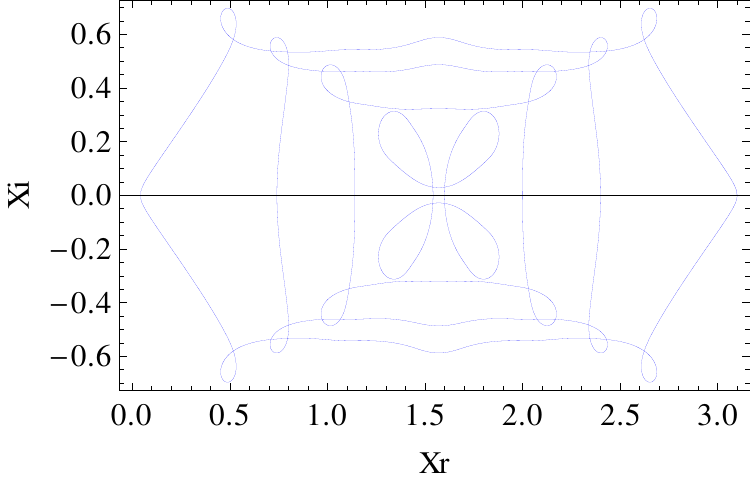}} 
\caption{ The numerically evaluated complex trajectories for the particle in  coherent state in an infinite well potential, in the $J=$ 0.04, 0.09, 0.16, 0.25, 0.36 cases. In all cases, trajectories are plotted for the initial values of $x_r$ =1.6, 2.0, 2.4 and 3.1, respectively}  \label{fig:infwel_coh02456}} 
 \end{figure}

As another example,  now consider a particle of mass $m$ and energy $E$, trapped in an infinite potential well of width $\pi a$. As  seen in Sec. 3, the classical trajectories are straight lines  in the complex plane, lying parallel to the real axis. Let us attempt to solve the quantum problem with a shifted Hamiltonian \cite{antoinepap}

\begin{equation}
H = -\frac{\hbar^2}{2m}\frac{d^2}{dx^2} -\frac{\hbar^2}{2ma^2},
\end{equation}
together with the boundary condition

\begin{equation}
\psi(x)=0, \qquad x\geq \pi a \; \; \; \hbox{and} \; \; \; x \leq 0.
\end{equation}
The normalized eigenstates and corresponding eigenvalues are

\begin{equation}
\psi_n(x) = \sqrt{\frac{2}{\pi a}} \sin (n+1) \frac{x}{a}, \qquad n=0,1, ...
\end{equation}
and

\begin{equation}
E_n=\hbar \omega e_n,
\end{equation}
with 

\begin{equation}
\omega = \frac{\hbar}{2ma^2} \qquad \hbox{and} \qquad e_n =n(n+2), \qquad n=0,1, ...
\end{equation}
With $a=1$, we have drawn the complex quantum trajectories corresponding to $n=0,1,2,3,4$ states of the infinite potential well eigenstates in Fig. (\ref{fig:infwel_qm01234}). The  coherent states of this particle can be written as \cite{antoinepap}
 
\begin{equation}
\phi_J (x,t)=\frac{1}{N(J)}\sum_{n\geq 0}\frac{J^{n/2}}{\sqrt{\frac{{n!} {(n+2)!}}{2}}}e^{-i n(n+2)\omega t}\psi_n(x). \label{eq:coh_infwel} 
\end{equation}
 The numerically evaluated complex trajectories for the particle in the $J=$ 0.04, 0.16, 0.25, 0.36 coherent states, with $\omega =1$, are shown in Fig. (\ref{fig:infwel_coh02456}). In this numerical approach, we have included terms up to $n=7$ in the expansion (\ref{eq:coh_infwel}). It is interesting to note the presence of almost straight trajectories parallel to the real axis, as in the case of classical complex trajectory solutions of free particles (Sec. 2).  But near the turning points, the trajectories are more like those near the turning points of classical symmetric Poschl-Teller potentials. In this case, deviations from the classically expected straight line trajectories (mentioned above) at the inner region do not seem to arise from  truncation errors in the numerical method. This can be seen by  varying the number of terms in the series (\ref{eq:coh_infwel}). The curves are the same even if we include  only terms up to $n=4$. But drawing trajectories for much larger values of $J$ than that given in Fig. (\ref{fig:infwel_coh02456}) is not possible with the present numerical method.   For such values of $J$, the trajectories  are not  confined to the physically limited interval of $0<x_r< \pi$ while using this method.

\begin{figure} [ht]
\centering{\resizebox {1 \textwidth} {0.2 \textheight }  
{\includegraphics {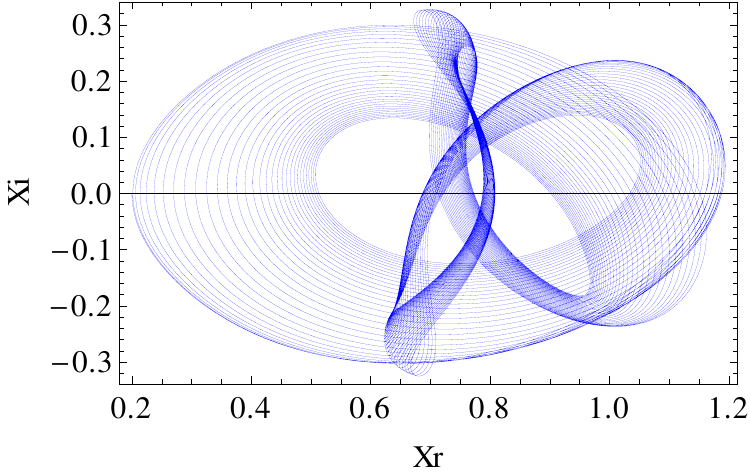} \includegraphics {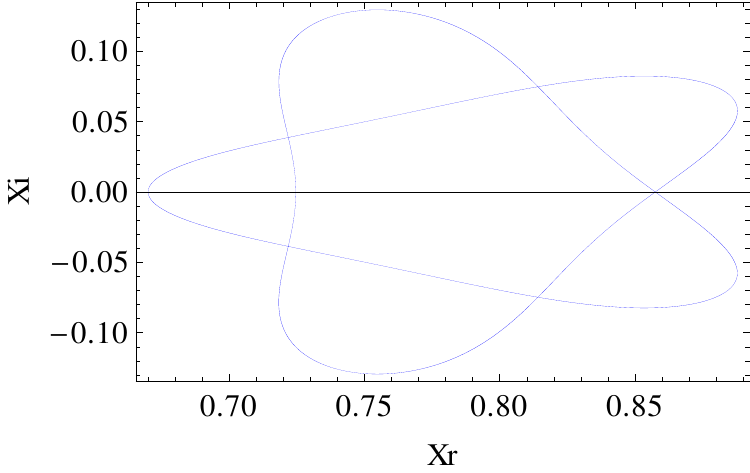} \includegraphics {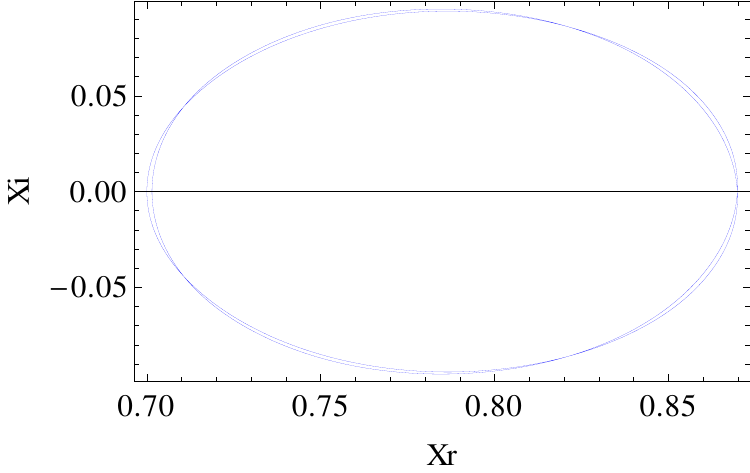} } 
\caption{ The complex trajectories for the particle in a Gazeau-Klauder coherent state with $J$ =0.16  in the Poschl-Teller potential. The trajectories in different plots are  for the initial values on the real line $x_{r0}$ = 0.2, 0.67, 0.7015 respectively and for $0<t<100$ in each case.}  \label{fig:PT_coh3}} 
 \end{figure}

The next problem  we consider is the symmetric Poschl-Teller potential

\begin{equation}
V(x) =2V_0 \frac {l (l -1)}{\sin^2(2x)},\label{eq:symPT}
\end{equation}
whose classical trajectories are discussed in Sec. 3. We have taken $l =k$ and $2a=1$ in Eq. (\ref{eq:PT_pot}). In addition, we consider a shifted Hamiltonian \cite{antoinepap}

\begin{equation}
H=-\frac{\hbar^2}{2m}\frac{d^2}{dx^2}-2\frac{\hbar^2}{m}l^2 + V(x).
\end{equation}
Solving the Schrodinger equation with the boundary condition $\psi (0)=\psi(\pi /2)=0$, the normalized eigenstates  can be found as

\begin{equation}
\psi_n(x) =[c_n(l)]^{-1/2} \cos^{l}(x)\sin^{l }(x) \; \mbox{$_2$F$_1$}(-n,n+2l;l +(1/2);\sin^2x). \label{eq:psi_PT}
\end{equation}
Here we choose $l =3/2$. The  Gazeau-Klauder coherent states are  \cite{antoinepap}
 
\begin{equation}
\phi_J(x,t) = \frac{\sqrt{3!}}{N(J)}\sum_{n\geq 0}\frac{J^{n/2}}{\sqrt{{{n!} {(n+3)!}}}}e^{-i n(n+3)\omega t}\psi_n(x), \label{eq:coh_PT} 
\end{equation}
where $\omega = {\hbar}/{2ma^2}$. The complex quantum trajectories, evaluated numerically with $\omega =1$, a value of $J=0.16$ and for different initial points are shown in Fig. (\ref{fig:PT_coh3}). Terms up to $n=4$ are included in the series.

These trajectories  are not identical to those of the corresponding classical case; i.e., there is no congruence with the classical trajectories in this case.  
However, some  interesting features of   these trajectories may  be noted. First,  for initial points close to the boundary  of the potential, they spiral inwards with a period;  i.e., the maximum displacements from the center to either side on the real and imaginary axes  decrease  with each cycle. This can be seen clearly from the first trajectory in Fig. (\ref{fig:PT_coh3}), which has starting point $x_{r0}=0.2$.   When the initial point is nearer to the bottom of the potential, such as $x_{r0}$ =  0.67 shown in the second figure, we have a pentagonal star-shaped trajectory.  The point $x_{r0}=0.7015$ is special for the case of $J=0.16$, since it leads  to an almost circular trajectory in the complex plane and the particle  remains in it for ever. It is interesting to note that if we follow the trajectory for large $t$, the pattern evolves  with time and passes through these different phases. Thus the trajectory in the third panel is the final one, in the limit $t\rightarrow \infty$, for whatever initial point one starts with. In particular, we have observed that the trajectory with the initial point $x_{r0}=0.2$, drawn in Fig. (\ref{fig:PT_coh3}), approaches this final shape asymptotically. During this process, also its orientation changed from anti-clockwise to clockwise and period changed from $\pi$ to $\pi/2$. Including only terms up to $n=3$ does not significantly affect the  shape of this or neighboring trajectories. Hence we conclude that also in Fig. (\ref{fig:PT_coh3}), the shape of the trajectories do not arise from truncation errors. For values of $J$ much larger than that in this figure, the trajectories cannot be drawn since they are not  confined to the physically limited interval of $0<x_r<\pi/2$. This may be due to the error in truncation.

There is another class of coherent states based on supersymmetric quantum mechanics (SUSYQM) \cite{bergeron1,bergeron2}. In the case of symmetric Poschl-Teller potential given in Eq. (\ref{eq:symPT}), let us take $V_0=\hbar^2/m$ and keep the Hamiltonian as 

\begin{equation}
H_l=-\frac{\hbar^2}{2m}\frac{d^2}{dx^2}+\frac{2\hbar^2}{m}\frac{l(l-1)}{\sin^2(2x)}. \label{eq:PT_hamilt}
\end{equation}
Then the energy eigenvalues can be written as

\begin{equation}
E_n=\frac{2\hbar^2}{m}(n+l)^2,
\end{equation}
while the eigenstates $\psi_n(x)$ are the same as that in Eq. (\ref{eq:psi_PT}).

\begin{figure} [ht]
\centering{\resizebox {1 \textwidth} {0.2 \textheight }  
{\includegraphics {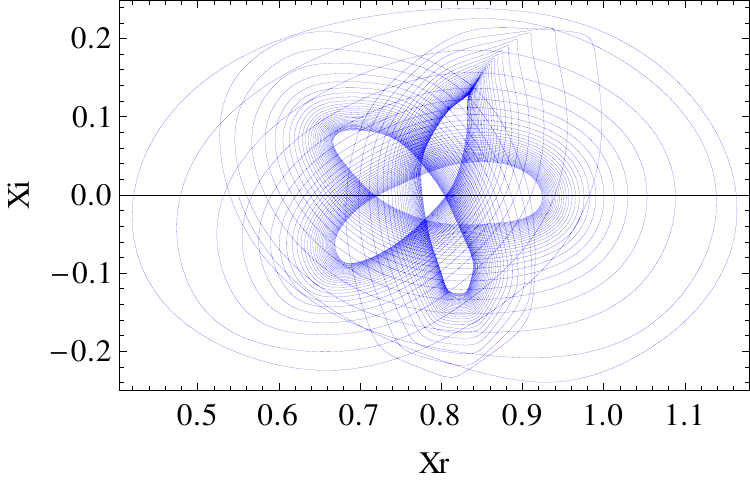} \includegraphics {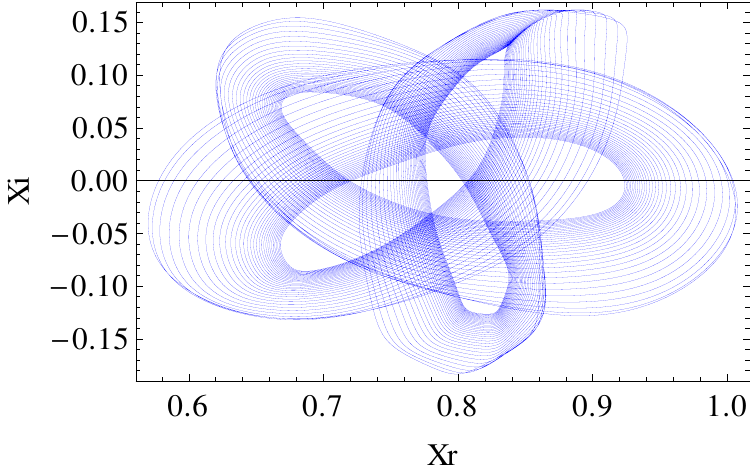} \includegraphics {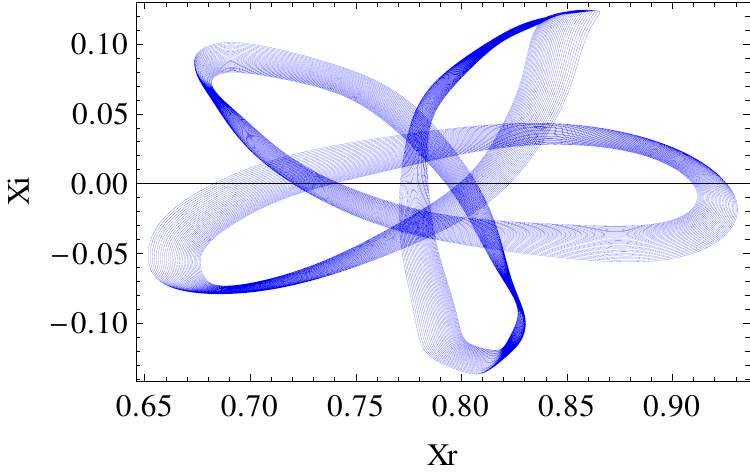} \includegraphics {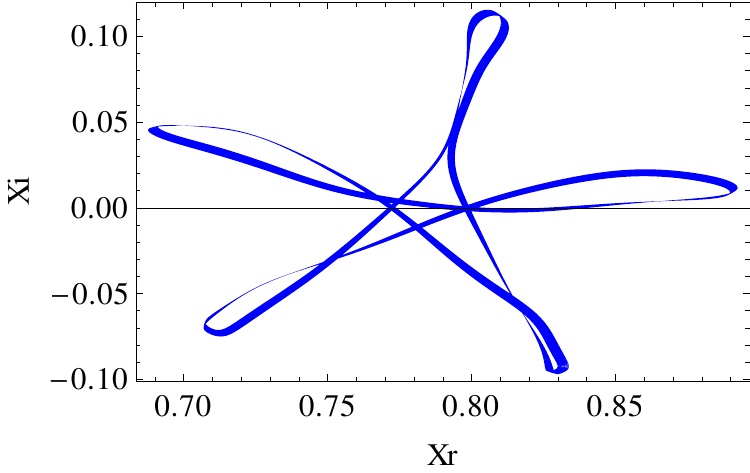}} 
\caption{ The complex trajectories for the particle in a  SUSYQM coherent state with $q =0.8$ and $k=0.5$  in the Poschl-Teller potential. The trajectories in the different plots are  for the initial values on the real line $x_{r0}$ = 0.55, 0.65, 0.72, $\pi/4$, respectively and for $0<t<100$ in each case.}  \label{fig:PT_SUSY_coh3}} 
 \end{figure}

The superpotential $W_l(x)$ in this case are defined as  $W_l(x)=-2\hbar l\cot(2x)$ \cite{bergeron1,bergeron2}
and the lowering and raising operators $A_l$ and $A_l^{\dagger}$ are, respectively,

\begin{equation}
A_l \equiv W_l +\hbar \frac{d}{dx} \qquad \hbox{and} \qquad A_l^{\dagger} \equiv W_l -\hbar \frac{d}{dx}.
\end{equation}
The Hamiltonian $H_l$  now becomes $H_l=\frac{1}{2m}A_l^{\dagger}A_l+E_0$. The SUSYQM coherent states are the normalized eigenvectors of $A_l$ and can be written as

\begin{equation}
\eta_{q,k}(x)=N(q)e^{(-3\cot(2q)+ik)x}\sin^{3/2}(2x), \label{eq:eta}
\end{equation}
where we have taken $l=3/2$ as in the previous case of Gazeau-Klauder coherent states. The corresponding eigenvalues are $-3\hbar \cot(2q)+i\hbar k$. The time evolution of this state can be found using the usual approach of expanding $\eta_{q,k}(x,t)$ in terms of the eigenfunctions $\psi_n(x)$ given in Eq. (\ref{eq:psi_PT}):

\begin{equation}
\eta_{q,k}(x,t)=\sum_{n=0}^{\infty}a_n \psi_n(x) e^{-i\frac{E_n}{\hbar}t}=\sum_{n=0}^{\infty}c_n \psi_n(x) e^{-i(n+\frac{3}{2})^2 \omega t} \label{eq:eta_xt}
\end{equation}
where $\omega =2\hbar/m$ in this case. Then the coefficients $a_n$ can be found as 

$$
a_n=\int_{0}^{\pi/2}\psi_n^{\star}(x) \eta_{q,k}(x,0)dx. 
$$
Putting $q=0.8$ and $k=0.5$, the values obtained numerically for the first few of  these coefficients are $a_0 = 0.80121555+ i\; 0.33440120$, $a_1=0.02393104- i\; 0.10634033$, $a_2=-0.00838162 -i\; 0.00039834$, $a_3=0.00168689- i\; 0.00677396$, $a_4=-0.00133694 -i\; 0.00007039$, $a_5=0.00041423 -i\;0.00164860$. Using these  in Eq. (\ref{eq:eta_xt}), complex quantum trajectories were drawn for various initial values, following the  numerical method adopted in the previous cases. Fig.  \ref{fig:PT_SUSY_coh3} shows such trajectories for  initial values $x_{r0}$ = 0.55, 0.65, 0.72, and $\pi/4$, respectively.  It was found that  the situation is similar to that of the  Gazeau-Klauder coherent states for the same potential, but has quite different patterns for the complex trajectories. Trajectories which start from points  away from the minimum of the potential ($x_{r} =\pi/4$) can be seen to inspiral as in the previous case,  but they appear to be more symmetrical (For example, see the first two diagrams in the panel of Fig.  \ref{fig:PT_SUSY_coh3}). Similarly for points close to the minimum of the potential (for instance, $x_{r0}=0.72$), the pattern evolves to that of a pentagonal flower. Moreover, the final temporally stable pattern (the last one  in  Fig.  \ref{fig:PT_SUSY_coh3}, where $x_{r0} =\pi/4$) is not a circle, which demonstrates that this SUSYQM coherent state is quite distinct from Gazeau-Klauder coherent states.

We have attempted to draw these trajectories for various combinations of the values of $q$, $k$ and starting points $x_{r0}$. In all cases, the shape of trajectories remain the same. However, it was not possible to draw the trajectories for values of both $q$ and starting points  near the turning points and $k>>1$, possibly due to truncation errors. In such cases, the trajectories were not found confined to the physically limited interval of $0<x_r<\pi/2$.

\section{Conclusion}

  Using the  MdBB  approach, we have illustrated     that the    coherent state trajectories of the harmonic oscillator are identical  to the complex classical  trajectories. This feature  is not restricted to high energies alone. We  note that  this result is not totally unexpected for standard coherent states because of their holomorphic nature. However, such congruence is as yet reported only in the  MdBB quantum trajectory representation.  The analysis  is extended to   potentials such as an infinite well and a symmetric Poschl-Teller potential.   We have used numerical methods for the plotting of complex trajectories in such cases. It is found that for a particle trapped in an infinite potential well, the coherent state  complex trajectories partly agree with the classical solutions. For large values of the parameter $J$, the agreement becomes almost perfect when initial points are located close to the turning points. On the other hand, for the Poschl-Teller potential, no congruence is observed and hence no exact classical correspondence can be claimed for both Gazeau-Klauder and SUSYQM coherent states. But we find that   the complex trajectories of  these coherent states become  periodic as $t \rightarrow \infty$, for typical values of the parameters involved. Most interestingly, different generalized coherent states have different complex trajectories with distinguishable patterns.

 There were several attempts to formulate complex quantum mechanics, such as that in \cite{rivers} and references therein, where complex space variables appear. But they are not directly related to the de Broglie-Bohm quantum trajectories approach and does not envisage individual particle trajectories as in Bohmian mechanics.  However, it remains a strong possibility that such works  on complex quantum mechanics shall be of help to explain, in a more comprehensive way, the physical interests behind the present work.  In this paper, our attempt is to see whether the complex classical trajectories of particles  \cite{classic1,classic2a,classic3} in various potentials  are identical to the complex trajectories in a modified version of de Broglie-Bohm  quantum theory, for the corresponding  coherent states. Since this is our main concern, other complex formulations of quantum mechanics  are not considered here. But we wish to highlight that to proceed further, the complex quantum mechanics developed in the above references can be of immense help.

Exploring the classical correspondence of coherent states at the level of quantum trajectories  is capable of revealing more information on the fundamental nature of  coherent states. It is interesting to note how the individual eigenstates which lead to trajectories in Fig.(\ref{fig:infwel_qm01234}) combine to form a coherent state with quantum trajectories shown in Fig. (\ref{fig:infwel_coh02456}), that at least partly agree with the classical solutions. Similarly, the eigenstates of Poschl-Teller potential combine to form coherent states having trajectories  that correspond to periodic classical trajectories.  In general,  such trajectories have  very interesting properties, which  makes the method a potential tool  in analyzing the  properties of generalized coherent states.

%\section{References}

\begin{acknowledgements}

MVJ wishes to thank Professors N. D. Hari Dass and M. Raveendranadhan for  discussions and the Chennai Mathematical Institute, Chennai, India for hospitality during a short visit.
\end{acknowledgements}

% BibTeX users please use one of
%\bibliographystyle{spbasic}      % basic style, author-year citations
%\bibliographystyle{spmpsci}      % mathematics and physical sciences
%\bibliographystyle{spphys}       % APS-like style for physics
%\bibliography{}   % name your BibTeX data base

\begin{thebibliography}{99}
 \bibitem{db1}  de Broglie, L.:  { Ph.D. Thesis.} (University of Paris) (1924)
 \bibitem{db2}  de Broglie, L.:  { J. Phys. Rad., 6$^e$ serie, t.}    {\bf 8}, 225 (1927)
\bibitem{valentini}  Bacciagaluppi, G.,  Valentini, A.:  { Quantum Theory at the Crossroads.}  Cambridge University Press, Cambridge (2009)
\bibitem{dBB}   Bohm, D.,  Hiley, B.J.:   { The Undivided
Universe.}  Routledge, London and New York (1993)
\bibitem{holland}  Holland, P.:   { The Quantum Theory of Motion.} Cambridge  University Press, Cambridge (1993)
\bibitem{carroll}  Carroll, R.:   { Quantum Theory, Deformation, and Integrability.} North Holland (2000)
\bibitem{wyattbook}  Wyatt, R.E.: {Quantum Dynamics with Trajectories: Introduction to Quantum Hydrodynamics.} Springer, New York (2005)
\bibitem{pratim}  Chattaraj, P.K.: (ed.)  { Quantum Trajectories.}  CRC Press, Taylor \& Francis Group, Boca Raton, London and New York (2011)
\bibitem{floyd}  Floyd, E.R.: Modified potential and Bohm's quantum-mechanical potential.  { Phys. Rev.} D {\bf 26}, 1339 (1982)
\bibitem{faraggi}  Faraggi, A.,   Matone. M.: Quantum mechanics from an equivalence principle.  {Phys. Lett.} B {\bf 450}, 34  (1999)
\bibitem{mvj1}  John, M.V.:  Modified de Broglie-Bohm approach to quantum mechanics. { Found. Phys. Lett.}  {\bf 15}, 329 (2002)
\bibitem{yang1}  Yang, C.-D.: Quantum dynamics of hydrogen atom  in complex space. { Ann. Phys. (N.Y.)} {\bf 319},  399 (2005)
\bibitem{yang2}  Yang, C.-D: Wave-particle duality in complex space. { Ann. Phys. (N.Y.)}  {\bf 319},  444 (2005)
\bibitem{yang3} Yang, C.-D: Modeling quantum harmonic oscillator in complex domain. { Chaos, Solitons Fractals} {\bf 30},  342 (2006)
\bibitem{goldfarb} Goldfarb, Y.,  Degani, I.,    Tannor, D.J.: Bohmian mechanics with complex action: A new trajectory-based formulation of quantum mechanics. { J. Chem. Phys.} {\bf 125},  231103 (2006)
\bibitem{wyatt1}  Chou, C.-C.,   Wyatt, R.E.: Computational method for the quantum Hamilton-Jacobi equation: One-dimensional scattering problems. { Phys. Rev.} E {\bf 74},  066702 (2006)
 \bibitem{wyatt2}  Chou, C.-C.,   Wyatt, R.E.: Computational method for the quantum Hamilton-Jacobi equation: Bound states in one-dimension. { J. Chem. Phys.} {\bf 125},  174103 (2007)
\bibitem{sanz1}  Sanz, A.S.,   Miret-Artes, S.: Aspects of nonlocality from a quantum trajectory perspective: A WKB approach to Bohmian mechanics. { Chem. Phys. Lett.} {\bf 445},  350 (2007)
\bibitem{sanz2}  Sanz, A.S.,  Miret-Artes, S.: Comment on "Bohmian mechanics with complex action: A new
trajectory-based formulation of quantum mechanics"
[J. Chem. Phys. 125, 231103, (2006)]. { J. Chem. Phys.} {\bf 127}, 197101 (2007)
\bibitem{goldfarb2}  Goldfarb, Y.,  Degani, I.,   Tannor, D.J.: Response to "Comment on 'Bohmian mechanics with complex action: A new trajectory-based formulation of quantum mechanics' "
[J. Chem. Phys. 127, 197101 (2007)].
 { J. Chem. Phys.} {\bf 127},  197102 (2007)

\bibitem{mvj2}  John, M.V.: Probability and complex quantum trajectories. { Ann. Phys.} {\bf 324}, 220 (2009)
\bibitem{mvj3} John, M.V.: Probability and complex quantum trajectories: Finding the missing links. { Ann. Phys.} {\bf 325}, 2132 (2010)

\bibitem{klauderbook} Klauder, J.R.,  Skagerstam, B.:  { Coherent States - Applications in Physics and Mathematical Physics.}  World Scientific, Singapore (1985)
\bibitem{antoinebook} Ali, S.T., Antoine, J.-P.,  Gazeau, J.-P.:  { Coherent States, Wavelets and Their Generalizations.}  Springer-Verlag, New York (2000)

\bibitem{classic1}  Bender, C.M.,  Boettcher, S.,   Meisinger, P.N.: PT-symmetric quantum mechanics. { J. Math. Phys.} {\bf 40}, 2201 (1999)
%\bibitem{classic2}  Nanayakkara, A.:  { Czech. J. Phys.} {\bf 54}, 101 (2004)
\bibitem{classic2a}  Nanayakkara, A.: Classical trajectories of 1D complex non-Hermitian
Hamiltonian systems.   { J. Phys. A: Math. Gen.} {\bf 37}, 4321 (2004)
\bibitem{classic3}  Bender, C.M., Chen, J.-H, Darg, D.W.,  Milton, K.A.: Classical trajectories for complex Hamiltonians. { J. Phys. A: Math. Gen.} {\bf 39}, 4219 (2006)
\bibitem{antoinepap} Antoine, J.-P., Gazeau, J.-P., Monceau, P., Klauder, J.R.,  Penson, K.A.: Temporally stable coherent states for infinite well and Poschl-Teller potentials. { J. Math. Phys.} {\bf 42}, 2349 (2001)
\bibitem{bergeron1} Bergeron, H., Gazeau, J.-P., Siegl, P., Youssef, A.: Semi-classical behavior of Poschl-Teller coherent states. {Eur. Phys. Lett.} {\bf 92}, 60003 (2010)
\bibitem{bergeron2} Bergeron, H.,  Siegl, P., Youssef, A.: New SUSYQM coherent states for Poschl-Teller potentials: a detailed mathematical analysis. {J. Phys. A: Math. Theor.} {\bf 454}, 244028 (2012)
\bibitem{rivers} Rivers, R.J.: Path Integrals for (Complex) Classical and Quantum Mechanics. {arXiv:1202.4117} (2012)

%
% and use \bibitem to create references. Consult the Instructions
% for authors for reference list style.
%
%\bibitem{RefJ}
% Format for Journal Reference
%Author, Article title, Journal, Volume, page numbers (year)
% Format for books
%\bibitem{RefB}
%Author, Book title, page numbers. Publisher, place (year)
% etc
\end{thebibliography}

% Non-BibTeX users please use

\end{document}